\documentclass[aps,twocolumn,prb,amsmath,amssymb,superscriptaddress,longbibliography]{revtex4-2}
\usepackage{url}
\usepackage{graphicx}
\usepackage{color}
\usepackage{natbib}
\usepackage{hyperref}
\usepackage{notes2bib}
\usepackage{mathbbol}
\usepackage{float}
\usepackage{tikz}
\usepackage{xcolor}
\usepackage{dsfont}
\usepackage{svg}
\usepackage[normalem]{ulem}

\begin{document}

\title{Quantum theory for edge current and noise in two-dimensional topological superconductors}
\author{S. Pintus}
\author{A. Cr\'epieux}
\affiliation{Aix Marseille Univ, Universit\'e de Toulon, CNRS, CPT, Marseille, France}
\date{\today}

\begin{abstract}
We calculate the edge current and its fluctuations, i.e. noise, in a two-dimensional topological superconductor using the T-matrix and the Green function techniques. We show that the current is zero for non-chiral edge states and non-zero for chiral edge states, while the edge noise is non-zero whatever the chirality of the edge states. By applying our results to toy models with chiral edge states, we find that the noise is closely related to the Chern number. The edge noise is non-zero only when the Chern number is non-zero, and the bulk noise exhibits a peak each time the Chern number varies, meaning that there are strong current fluctuations when a topological phase transition occurs. Our results suggest that the bulk noise could be seen as a topological susceptibility. In the extended Qi-Wu-Zhang model, where an edge state is present even though the Chern number is zero, the edge noise no longer follows the same behavior as it does in the non-zero Chern number model. Instead, it is related to other topological invariant, such as the Zak phase.
\end{abstract}


\maketitle


\section{Introduction}

Topological superconductors are the subject of an increasing number of theoretical and experimental studies. In addition to exhibiting exotic properties with potential applications\cite{Qi2011,Bernevig2013,Beenakker2016,Sato2017,Sharma2022,Mandal2023}, the most fascinating aspect is that their physical characteristics differ significantly from those of non-topological superconductors. The most notable are the fact that a phase transition is not necessarily accompanied by symmetry breaking\cite{Berezinskii1970,Berezinskii1972,Kosterlitz1973}, that a gap closing phenomenon in the spectral function may occur at the topological phase transition point\cite{Tanaka2012,Ezawa2013,Menard2017,Crepieux2023}, that topologically protected chiral edge states allowing current to flow without dissipation may exist\cite{Ying2022,Zhang2025} and can host Majorana modes\cite{Kallin2016}.

Although the study of edge states has progressed significantly, a relatively limited number of theoretical works devoted to the study of edge current and noise in topological superconductors exist. Concerning the edge current, most of these works used quasi-classical approaches\cite{Stone2004,Huang2014,Huang2015,Suzuki2016,Wang2018,Nie2020,Holmvall2023}, such as the Ginzburg-Landau theory or the Eilenberger equation. A few others used the Green function formalism\cite{Braunecker2005,Schnyder2013,Holst2022,Pathak2024}. Noise has been theoretically studied in topological superconducting wires\cite{Akhmerov2011,Diez2014,Valentini2016} and in normal metal-topological superconductor junctions\cite{Bathellier2019,Jonckheere2020,Kokkeler2025}, showing evidence of Majorana fermions. The cross-correlator of the currents was calculated  in topological superconductor beam splitters\cite{Haim2015,Jonckheere2017}
and heterostructures made of nanowire, superconductor and altermagnet\cite{Mondal2025}, and it was shown that its sign changes when a transition from a topological superconductor to a conventional one occurs. However, the calculation of edge noise in two-dimensional topological superconductors taking into account the crystal lattice structure of the sample has not yet been carried out, except in one pioneering work\cite{Gnezdilov2015}.

A quantum theory to calculate edge current and edge noise in a unified way is thus highly necessary and has to be developed without further delay, especially since there are now experimental techniques for measuring differential resistance in Weyl superconductors\cite{Wang2020} and for imaging edge current\cite{Uri2020}. This article aims to address this issue. It is organized as follows: the method used to calculate the edge spectral function, the edge current, and the edge noise in a two-dimensional topological superconductor is presented in Section~II, where the expressions for these quantities are given in terms of the retarded Green function for the first one, and in terms of Keldysh Green function for the latter two. We apply the results to various 2D systems in Section~III, and we conclude in Section~IV.


\section{Method and results}

\begin{figure}[b]
    \centering
    \includegraphics[width=0.8\linewidth]{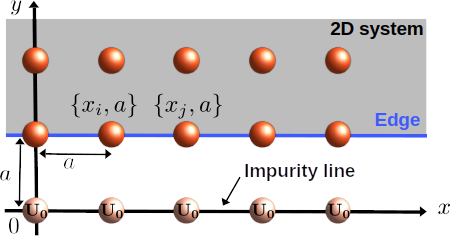}
    \caption{Schematic picture of the 2D system, modeled as a square lattice, with an impurity line at $y=0$ with significant strong potential $U_0$, and an edge at $y=a$, where $a$ is the inter-atomic distance. The translation invariance symmetry is broken along $y$-axis while it remains preserved along the $x$-axis.}
    \label{fig:dispositif}
\end{figure}

In this Section, we give the details of the method used to calculate the spectral function, current, and noise along the edge located at position $y=a$ in the semi-infinite 2D system pictured in Fig.~\ref{fig:dispositif}. It is based on the determination of the Green function of a semi-infinite plane, calculated using the T-matrix\cite{Pinon2020,Pinon2021,Crepieux2023,Braz2024,Xiong2024}. In that respect, it differs from the other methods used to determinate the energy bands at the edge of the 2D system, based on the calculation of the eigenvalues of a finite-size system Hamiltonian\cite{QWZ2006,Yan2018}, or of the spectral function of a periodic long strip\cite{Pathak2024}. The advantage of our method is that it allows us to express the edge spectral function, the edge current, and the noise using a single quantity related to the Green function of the semi-infinite plane.

\subsection{Edge spectral function}

The edge spectral function $A_{k_xy=a}(\varepsilon)$ is obtained from the spectral function defined as
\begin{eqnarray}\label{def_spectral_function}
 A_{k_xy}(\varepsilon)= \mathrm{Tr}[ \mathcal{A}_{k_xy;k_x y}(\varepsilon)]
\end{eqnarray}
where $\text{Tr}[~]$ is the trace, and the spectral matrix is defined as the imaginary part of the retarded Green function, i.e.,
\begin{eqnarray}\label{def_spectral}
   \mathcal{A}_{k_xy;k_x' y'}(\varepsilon)=-\frac{1}{\pi}\text{Im}\{{\bf G}^r_{k_xy;k_x'y'}(\varepsilon)\}
\end{eqnarray}

In Eq.~(\ref{def_spectral}) appears the partial Fourier transform of the retarded Green function, which is defined as
\begin{eqnarray}\label{retarded_green}
    {\bf G}^r_{k_xy;k_x'y'}(\varepsilon)&=&\int_{1BZ}\frac{dk_y}{2\pi}\int_{1BZ}\frac{dk_y'}{2\pi}\nonumber\\
    && \times e^{i(yk_y-y'k_y')}{\bf G}^r_{k_xk_y;k_x'k_y'}(\varepsilon)
\end{eqnarray}
where $1BZ$ corresponds to the first Brillouin zone. In the formalism of the T-matrix, one has
\begin{eqnarray}
{\bf G}^r_{k_xk_y;k_x'k_y'}(\varepsilon)&=&\delta(k_x-k_x')\delta(k_y-k_y'){\bf g}^r_{k_xk_y}(\varepsilon)\nonumber\\
&&+{\bf g}^r_{k_xk_y}(\varepsilon){\bf T}_{k_xk_y;k_x'k_y'}(\varepsilon){\bf g}^r_{k_x'k_y'}(\varepsilon)
\end{eqnarray}
where ${\bf T}_{k_xk_y;k_x'k_y'}(\varepsilon)$ is the T-matrix. The bare Green function is given by ${\bf g}^r_{k_xk_y}(\varepsilon)=(\varepsilon+i\eta-\mathcal{H}_{k_xk_y})^{-1}$, where $\mathcal{H}_{k_xk_y}$ is the Hamiltonian of the infinite clean 2D system, and $\eta\rightarrow 0^+$ is the damping. The edge located at position $y=a$ is modeled by an impurity line of strong potential amplitude located at position $y=0$ (see Fig.~\ref{fig:dispositif}), with $a$ being the inter-atomic distance. In that case, the associated T-matrix ${\bf T}_{k_xk_y;k_x'k_y'}(\varepsilon)\equiv \delta(k_x-k_x'){\bf T}_{k_x}(\varepsilon)$ does not depend on $k_y$ or $k_y'$, and one has\cite{Pinon2020,Pinon2021}
\begin{eqnarray}\label{T_matrix}
    {\bf T}_{k_x}(\varepsilon)=\left[{\bf I}-{\bf U}\int_{1BZ}\frac{dk_y}{2\pi} {\bf g}^r_{k_xk_y}(\varepsilon) \right]^{-1}{\bf U}
\end{eqnarray}
with ${\bf I}$ being the identity matrix, and ${\bf U}$ being the impurity matrix. It leads to
\begin{eqnarray}
{\bf G}^r_{k_xk_y;k_x'k_y'}(\varepsilon)&=&\delta(k_x-k_x')\delta(k_y-k_y'){\bf g}^r_{k_xk_y}(\varepsilon)\nonumber\\
&+&\delta(k_x-k_x'){\bf g}^r_{k_xk_y}(\varepsilon){\bf T}_{k_x}(\varepsilon){\bf g}^r_{k_xk_y'}(\varepsilon)
\end{eqnarray}
By inserting this expression in Eq.~(\ref{retarded_green}), one gets for the retarded Green function
\begin{eqnarray}\label{edge_green}
    {\bf G}^r_{k_xy;k_x'y'}(\varepsilon)&=&\delta(k_x-k_x')\Big[{\bf g}^r_{k_x0}(\varepsilon)\nonumber\\
    &&+{\bf g}^r_{k_xy}(\varepsilon){\bf T}_{k_x}(\varepsilon){\bf g}^r_{k_x-y'}(\varepsilon)\Big]
\end{eqnarray}
where ${\bf g}^r_{k_xy}(\varepsilon)=\int_{1BZ}\frac{dk_y}{2\pi}e^{iyk_y}{\bf g}^r_{k_xk_y}(\varepsilon)$ is the partial Fourier transform of the bare Green function.


\subsection{Edge current}

The current operator between first neighbors at positions $\{x_i,y\}$ and $\{x_j,y\}$ (see Fig.~\ref{fig:dispositif}) at time $\tau$ is defined as
\begin{eqnarray}\label{def_current}
    \widehat I_{\langle i,j\rangle}(\tau)= -\frac{eit}{\hbar}\sum_{\sigma}\widehat c^{\,\dagger}_{x_i y \sigma}(\tau)c_{x_j y \sigma}(\tau)+h.c.
\end{eqnarray}
where $t $ is the hopping integral between first neighbors, $c^{\,\dagger}_{x_i y \sigma}$ and $c_{x_j y \sigma}$ are the creation and annihilation operators at positions $\{x_i,y\}$ and $\{x_j,y\}$, and $\sigma$ is the spin and/or orbital degree of freedom. Since the translation invariance is broken in the $y$ direction due to the presence of an impurity line at position $y=0$, we apply a partial Fourier transform by taking
\begin{eqnarray}
    \widehat c_{x y \sigma}(\tau)=\frac{1}{\sqrt{N_x}}\sum_{k_x}e^{i k_x x}\,\widehat c_{k_x y \sigma}(\tau)
\end{eqnarray}
Thus,
\begin{eqnarray}
    \widehat I_{\langle i,j\rangle}(\tau)&=& -\frac{eit }{\hbar N_x}\sum_{\sigma}\sum_{k_x k_x'}e^{-i k_x x_i}e^{i k_x' x_j}\nonumber\\
    &&\times\widehat c^{\,\dagger}_{k_x y \sigma}(\tau)\widehat c_{k_x' y \sigma}(\tau)+h.c.
\end{eqnarray}

By using the fact that the current does not depend on position $x_i$ due to current conservation, the quantity of interest is the position-averaged current along the $x$-axis: $  I_y(\tau) = \sum_{x_i} \widehat I_{\langle i,j\rangle}(\tau) / N_x$, that is
\begin{eqnarray}
    I_y(\tau) &=& -\frac{eit }{\hbar N_x^2}\sum_{x_i\sigma}\sum_{k_x k_x'}e^{-i k_x x_i}e^{i k_x' x_j}\nonumber\\
    &&\times\widehat c^{\,\dagger}_{k_x y \sigma}(\tau)\widehat c_{k_x' y \sigma}(\tau)+h.c.
\end{eqnarray}
The positions $x_i$ and $x_j$ are first neighbors, thus one has $x_j=x_i+a$. By calculating the sum on $x_i$, we get
\begin{eqnarray}\label{Ipositionaverage}
    I_y(\tau) &=& -\frac{eit }{\hbar N_x}\sum_{\sigma}\sum_{k_x}e^{ia k_x}\widehat c^{\,\dagger}_{k_x y \sigma}(\tau)\widehat c_{k_x y \sigma}(\tau)+h.c.\nonumber\\
\end{eqnarray}
By introducing the non-equilibrium Green function defined as $G^<_{k_x y \sigma; k_x'y'\sigma'}(\tau,\tau')=i\langle \widehat c^{\,\dagger}_{k_x' y' \sigma'}(\tau')\widehat c_{k_x y \sigma}(\tau)\rangle $, one obtains for the current expectation value the expression
\begin{eqnarray}
   \langle I_y(\tau)\rangle= -\frac{2et }{\hbar N_x}\sum_{k_x}\text{Tr}\left[\text{Re}\left\{e^{ia k_x}{\bf G}^<_{k_xy;k_xy}(\tau,\tau)\right\}\right]\nonumber\\
\end{eqnarray}
where ${\bf G}^<_{k_xy;k_xy}(\tau,\tau')$ is the non-equilibrium Green function matrix whose elements are $G^<_{k_x y \sigma; k_x'y'\sigma'}(\tau,\tau')$. In the absence of time-dependent excitation, one has
\begin{eqnarray}
    {\bf G}^<_{k_xy;k_xy}(\tau,\tau')=\int_{-\infty}^\infty\frac{d\varepsilon}{2\pi}e^{i\varepsilon(\tau-\tau')/\hbar}{\bf G}^<_{k_xy;k_xy}(\varepsilon)\nonumber\\
\end{eqnarray}
In that case, the current is time-independent and reads
\begin{eqnarray}\label{result_current}
   \langle I_y\rangle&=& -\frac{2et }{\hbar}\int_{1BZ}\frac{dk_x}{2\pi}\int_{-\infty}^\infty \frac{d\varepsilon}{2\pi}\nonumber\\
   &&\times\text{Tr}\left[\text{Re}\left\{e^{ia k_x}{\bf G}^<_{k_xy;k_xy}(\varepsilon)\right\}\right]
\end{eqnarray}
This result allows us to re-derive the result obtained by  Pathak {\it et al.}\cite{Pathak2024} in the equilibrium limit. Indeed, one has  ${\bf G}^<_{k_xy;k_xy}(\varepsilon)=if(\varepsilon) \mathcal{A}_{k_xy;k_xy}(\varepsilon)$ in equilibrium\cite{Jauho2008}, where $f(\varepsilon)$ is the Fermi-Dirac distribution function, and $ \mathcal{A}_{k_xy;k_xy}(\varepsilon)$ the spectral matrix defined in Eq.~(\ref{def_spectral}). It leads to
\begin{eqnarray}\label{result_current_pathak}
   \langle  I_y\rangle&=& \frac{2et }{\hbar}\int_{1BZ}\frac{dk_x}{2\pi}\sin(k_xa)\nonumber\\
   &&\times \int_{-\infty}^\infty \frac{d\varepsilon}{2\pi}f(\varepsilon)\text{Tr}\left[ \mathcal{A}_{k_xy;k_xy}(\varepsilon)\right]
\end{eqnarray}
We note that if the spectral function is an even function with $k_x$, the current cancels out due to the presence of a sine factor in Eq.~(\ref{result_current_pathak}). This is the case for non-chiral edge states. It means that the current along the edge, corresponding to $\langle  I_{y=a}\rangle$, can be non-zero only when the edge states are chiral.


\subsection{Edge noise}

The finite-frequency noise is defined as the Fourier transform of the position-averaged current-current correlator
\begin{eqnarray}\label{def_noise}
\mathcal{S}_{yy'}(\omega)=\int_{-\infty}^{\infty} \langle \Delta I_y(\tau) \Delta I_{y'}(0) \rangle e^{-i\omega \tau}d\tau
\end{eqnarray}
with $ \Delta I_y(\tau)= I_y(\tau)-\langle I_y(\tau)\rangle$, where $ I_y(\tau)$ is given by Eq.~(\ref{Ipositionaverage}). We restricted our study to time-independent excitation. In that case $\langle I_y(\tau)\rangle=\langle I_y\rangle$, where $ \langle I_y\rangle$ is given by Eq.~(\ref{result_current}). We start by calculating the correlator $\mathcal{S}_{yy'}(\tau,\tau')= \langle \Delta I_y(\tau) \Delta I_{y'}(\tau') \rangle$. By using Eq.~(\ref{Ipositionaverage}), and assuming real value for the hopping integral $t$, we obtain
\begin{eqnarray}\label{noise}
 &&\mathcal{S}_{yy'}(\tau,\tau')=-\frac{4e^2t ^2}{\hbar^2 N_x^2}
 \sum_{\sigma\sigma'}\sum_{k_x k_x'}\sin(k_x a)\sin(k_x'a)\nonumber\\
&&\times
\langle\widehat c^{\,\dagger}_{k_x y \sigma}(\tau)\widehat c_{k_x y \sigma}(\tau)\widehat c^{\,\dagger}_{k_x' y' \sigma'}(\tau')\widehat c_{k_x' y' \sigma'}(\tau')\rangle
-\langle I_y\rangle^2\nonumber\\
\end{eqnarray}
By applying Wick theorem, we have
\begin{eqnarray}\label{wick}
    &&\langle\widehat c^{\,\dagger}_{k_x y \sigma}(\tau)\widehat c_{k_x y \sigma}(\tau)\widehat c^{\,\dagger}_{k_x' y' \sigma'}(\tau')\widehat c_{k_x' y' \sigma'}(\tau')\rangle\nonumber\\
    &&=\langle\widehat c^{\,\dagger}_{k_x y \sigma}(\tau)\widehat c_{k_x y \sigma}(\tau)\rangle\langle\widehat c^{\,\dagger}_{k_x' y' \sigma'}(\tau')\widehat c_{k_x' y' \sigma'}(\tau')\rangle
    \nonumber\\
    &&-\langle\widehat c^{\,\dagger}_{k_x y \sigma}(\tau)\widehat c_{k_x' y' \sigma'}(\tau')\rangle\langle\widehat c^{\,\dagger}_{k_x' y' \sigma'}(\tau')\widehat c_{k_xy \sigma}(\tau)\rangle
\end{eqnarray}
The first contribution on the right-hand side of Eq.~(\ref{wick}) will be canceled by the last term in Eq.~(\ref{noise}). Finally,
\begin{eqnarray}
 &&\mathcal{S}_{yy'}(\tau,\tau')=\frac{4e^2t ^2}{\hbar^2 N_x^2}
 \sum_{\sigma\sigma'}\sum_{k_x k_x'}\sin(k_x a)\sin(k_x'a)\nonumber\\
&&\times
\langle\widehat c^{\,\dagger}_{k_x y \sigma}(\tau)\widehat c_{k_x' y' \sigma'}(\tau')\rangle\langle\widehat c^{\,\dagger}_{k_x' y' \sigma'}(\tau')\widehat c_{k_xy \sigma}(\tau)\rangle
\end{eqnarray}
which leads to
\begin{eqnarray}
 &&\mathcal{S}_{yy'}(\tau,\tau')=\frac{4e^2t ^2}{\hbar^2 N_x^2}
 \sum_{\sigma\sigma'}\sum_{k_x k_x'}\sin(k_x a)\sin(k_x'a)\nonumber\\
&&\times
G^<_{k_x' y' \sigma'; k_xy\sigma}(\tau',\tau)G^>_{k_x y \sigma; k_x'y'\sigma'}(\tau,\tau')
\end{eqnarray}
By inserting this expression in Eq.~(\ref{def_noise}) and by performing the integration on time variable $\tau$, we obtain
\begin{eqnarray}
 &&\mathcal{S}_{yy'}(\omega)=\frac{4e^2t ^2}{\hbar}
\int_{1BZ}\frac{dk_x}{2\pi}\int_{1BZ}\frac{dk_x'}{2\pi}\sin(k_x a)\sin(k_x'a)\nonumber\\
&&\times \int_{-\infty}^\infty \frac{d\varepsilon}{2\pi}
\text{Tr}\Big[{\bf G}^<_{k_x' y'; k_xy}(\varepsilon){\bf G}^>_{k_x y ; k_x'y'}(\varepsilon-\hbar\omega)\Big]
\end{eqnarray}
Due to the presence of a factor $\delta(k_x-k_x')$ in the expression of the Green function (see Eq.~(\ref{edge_green})), the noise simplifies to
\begin{eqnarray}\label{result_noise}
 &&\mathcal{S}_{yy'}(\omega)=\frac{4e^2t ^2}{\hbar}
\int_{1BZ}\frac{dk_x}{2\pi}\sin^2(k_x a)\nonumber\\
&&\times \int_{-\infty}^\infty \frac{d\varepsilon}{2\pi}
\text{Tr}\Big[{\bf G}^<_{k_x y'; k_xy}(\varepsilon){\bf G}^>_{k_x y ; k_xy'}(\varepsilon-\hbar\omega)\Big]
\end{eqnarray}
In the equilibrium limit, by knowing that ${\bf G}^<_{k_xy';k_xy}(\varepsilon)=if(\varepsilon) \mathcal{A}_{k_xy';k_xy}(\varepsilon)$ and ${\bf G}^>_{k_xy;k_xy'}(\varepsilon-\hbar\omega)=-i(1-f(\varepsilon-\hbar\omega)) \mathcal{A}^\dag_{k_xy;k_xy'}(\varepsilon-\hbar\omega)$\cite{Jauho2008}, we obtain
\begin{eqnarray}\label{result_noise_equilibrium}
 \mathcal{S}_{yy'}(\omega)&=&\frac{4e^2t ^2}{\hbar}
\int_{1BZ}\frac{dk_x}{2\pi}\sin^2(k_x a)\nonumber\\
&&\times\int_{-\infty}^\infty \frac{d\varepsilon}{2\pi}
f(\varepsilon)(1-f(\varepsilon-\hbar\omega))\nonumber\\
&&\times 
\text{Tr}\Big[ \mathcal{A}_{k_x y'; k_xy}(\varepsilon) \mathcal{A}^\dag_{k_x y ; k_xy'}(\varepsilon-\hbar\omega)\Big]
\end{eqnarray}
This expression gives the bulk noise, $\mathcal{S}_\text{bulk}(\omega)\equiv\mathcal{S}_{yy}(\omega)$ with $y\gg a$, and the edge noise, $\mathcal{S}_\text{edge}(\omega)\equiv\mathcal{S}_{aa}(\omega)$, at equilibrium under the form of an integral on both energy $\varepsilon$ and $k_x$-momentum that can be computed numerically. Contrary to what was obtained for the edge current, i.e., only chiral edge states will have a non-zero edge current, one can have non-zero noise for both chiral and non-chiral edge states. Indeed, a factor $\sin^2(k_x a)$ is present in the noise expression given by Eq.~(\ref{result_noise_equilibrium}), while current exhibits a factor $\sin(k_x a)$ in the expression given by Eq.~(\ref{result_current_pathak}).


\section{Application}\label{application}

In this section, we use Eqs.~(\ref{result_current_pathak}) and (\ref{result_noise_equilibrium}) to numerically calculate the current~$\langle I_y\rangle$ and  noise~$\mathcal{S}_{yy}(\omega)$ in two 2D systems with chiral edge states, respectively described by the Qi-Wu-Zhang model and the bilayer model, with the objective being to characterize the similarities, if any, or the differences between these quantities. The Chern number is also calculated using\cite{Ghosh2010,Sedlmayr2017}
\begin{eqnarray}\label{chern_def}
 \mathcal{C}&=&\frac{i}{8\pi^2}\int_{1BZ} \text{d}{k_x}\int_{1BZ} \text{d}{k_y}\int_{-\infty}^\infty \text{d}\varepsilon\; \nonumber\\
 &&\times\text{Tr}\bigg[\mathcal{G}^2_{k_xk_y}(\varepsilon)(\partial_{k_y}\mathcal{H}_{k_xk_y})\mathcal{G}_{k_xk_y}(\varepsilon)(\partial_{k_x}\mathcal{H}_{k_xk_y})\nonumber\\
 &&-\mathcal{G}^2_{k_xk_y}(\varepsilon)(\partial_{k_x}\mathcal{H}_{k_xk_y})\mathcal{G}_{k_xk_y}(\varepsilon)(\partial_{k_y}\mathcal{H}_{k_xk_y})\bigg]
\end{eqnarray}
where $\mathcal{G}_{k_xk_y}(\varepsilon)=(i\varepsilon-\mathcal{H}_{k_xk_y})^{-1}$ is the Matsubara Green function. Moreover, we calculate the bulk and edge densities of states (DOSs), defined by
\begin{eqnarray}\label{DOS_def}
\rho_\text{bulk}(\varepsilon)&=&-\frac{1}{\pi}\iint_{1BZ} \text{d}{k_x} \text{d}{k_y}\mathrm{Im}\left\{\mathrm{Tr}\left[{\bf g}^r_{k_xk_y}(\varepsilon)\right]\right\}\\
\rho_\text{edge}(\varepsilon)&=&-\frac{1}{\pi}\int_{1BZ} \text{d}{k_x}\mathrm{Im}\left\{\mathrm{Tr}\left[{\bf G}^r_{k_xa;k_xa}(\varepsilon)\right]\right\}
\end{eqnarray}
respectively, where ${\bf g}^r_{k_xk_y}(\varepsilon)$ is the retarded bare Green function, and ${\bf G}^r_{k_xa;k_xa}(\varepsilon)$ is calculated from Eq.~(\ref{edge_green}).

In the following, we assume that the temperature at which we calculate the current and the noise is much lower than the superconductor critical temperature, and we neglect the temperature dependence in the superconducting order parameter.


\begin{figure}[t]
\begin{tikzpicture}
\node at (0,0) {
    \includegraphics[height=0.44\linewidth]{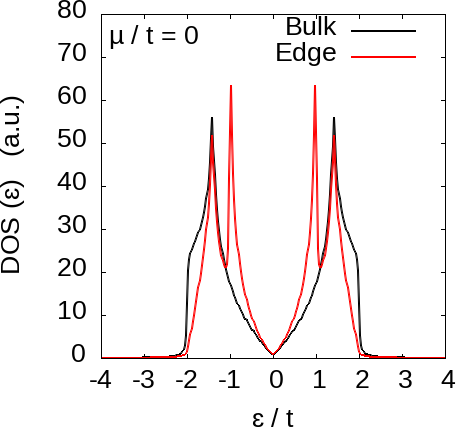}
    \includegraphics[height=0.44\linewidth]{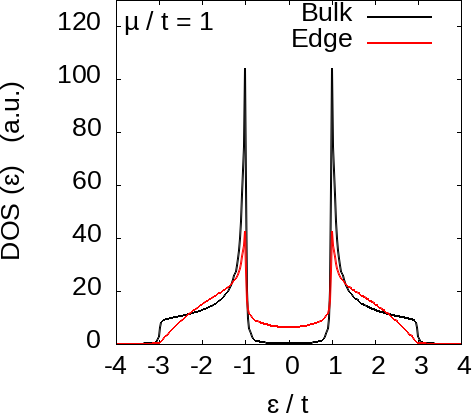}
    };
\node at (-3.9,1.6) {(a)};
\node at (0.3,1.6) {(b)};
\end{tikzpicture}
\begin{tikzpicture}
\node at (0,0) {   
    \includegraphics[height=0.44\linewidth]{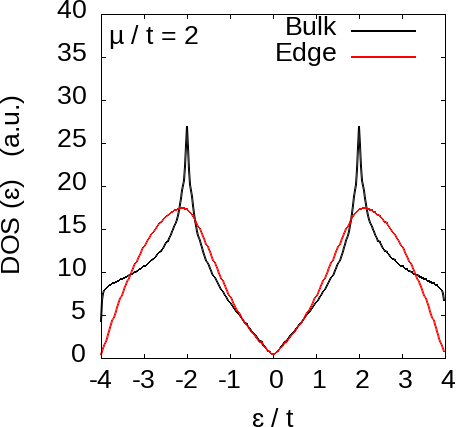}
    \includegraphics[height=0.44\linewidth]{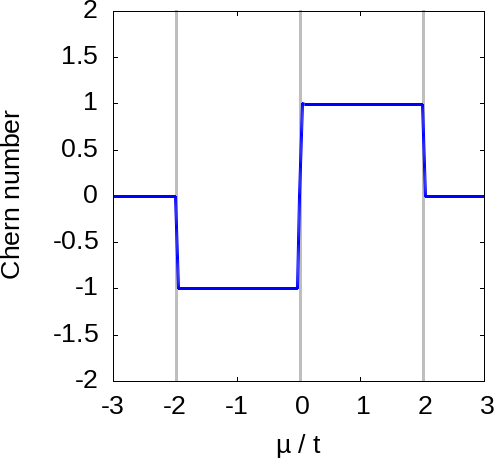}
    };
\node at (-3.9,1.6) {(c)};
\node at (0.3,1.6) {(d)};
\end{tikzpicture}
    \caption{Bulk and edge DOS at (a)~$\mu=0$, (b)~$\mu=t$, (c)~$\mu=2t$, and (d) Chern number $\mathcal{C}$ as a function of $\mu$ in the QWZ model. The parameters are $\Delta=t$, $\eta= 0.02t$ and $U_0=1000t$.  In panel (d), the vertical gray lines indicate the values of $\mu$ for which the Chern number value changes, i.e. $\mu=0$ and $\mu=\pm 2\,t$.}
    \label{fig:QWZmodel}
\end{figure}

\begin{figure}[t]
    \centering
    \includegraphics[width=0.32\linewidth]{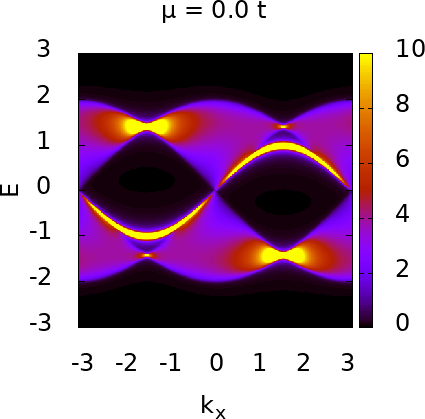}
    \includegraphics[width=0.32\linewidth]{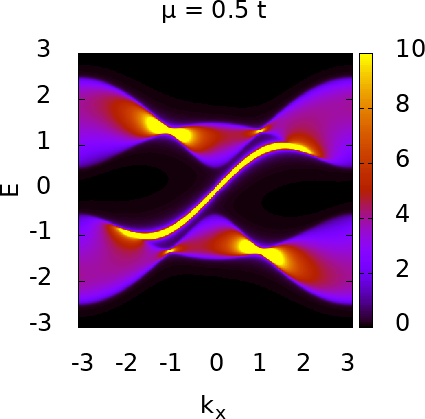} 
    \includegraphics[width=0.32\linewidth]{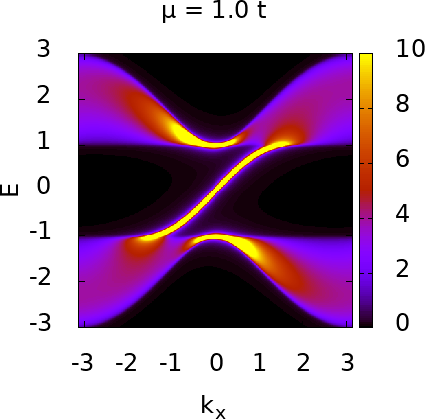}  \\
    \vspace{0.2cm}
    \includegraphics[width=0.32\linewidth]{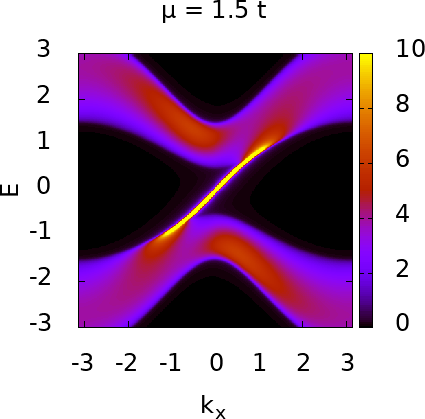}
    \includegraphics[width=0.32\linewidth]{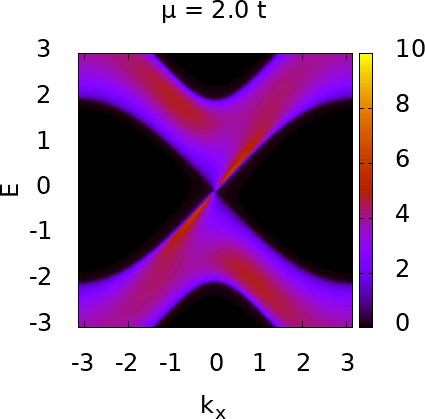}  
    \includegraphics[width=0.32\linewidth]{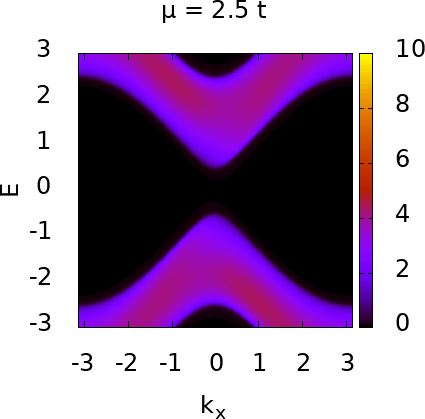}
   \caption{Edge spectral function $A_{k_xa}(\varepsilon)$ as a function of $k_x$ and $\varepsilon$ in the QWZ model for various values of $\mu$, with $\Delta=t$, $\eta= 0.02t$ and $U_0=1000t$.}
    \label{fig:QWZspectral1}
\end{figure}

\subsection{Qi-Wu-Zhang model}

We start with one of the simplest 2D model used to describe chiral topological superconductors, that is the Qi-Wu-Zhang (QWZ) model\cite{QWZ2006} given, in the basis $\{\widehat c_{k_xk_y},\widehat c^{\,\dagger}_{-k_x-k_y}\}$, by the $2\times 2$ matrix Hamiltonian
\begin{eqnarray}
    \mathcal{H}_{k_xk_y}=
    \begin{pmatrix}
    \mathcal{E}_{k_xk_y}& \mathcal{F}_{k_xk_y}\\
    \mathcal{F}^*_{k_xk_y}&-\mathcal{E}_{k_xk_y}\\
    \end{pmatrix}
\end{eqnarray}
with $\mathcal{E}_{k_xk_y}=\mu-t(\cos(ak_x)+\cos(ak_y))$ and $ \mathcal{F}_{k_xk_y}=\Delta(\sin(ak_x)-i\sin(ak_y))$. $\Delta$ is the superconducting order parameter, and $\mu$ the chemical potential. The QWZ eigenenergies for the 2D clean system read
\begin{eqnarray}
    E_\pm(k_x,k_y)=\pm \sqrt{\mathcal{E}_{k_xk_y}^{\,2}+|\mathcal{F}_{k_xk_y}|^2}
\end{eqnarray}   
The impurity matrix~${\bf U}$ used in the T-matrix formalism to model the impurity line is taken as
\begin{eqnarray}
   {\bf U}= \begin{pmatrix}
        U_0 & 0\\
        0 & -U_0
    \end{pmatrix}
\end{eqnarray}
where $U_0$ is the impurity strength. The DOS associated with the QWZ model is calculated in the absence (bulk DOS) and in the presence (edge DOS) of an impurity line located at $y=0$. It is displayed in Figs.~\ref{fig:QWZmodel}(a), (b) and (c), for three different values of $\mu$. For $\mu=t$, one observes a gap of energy in the bulk DOS and the presence of edge states inside the gap, called ingap states, with a non-zero edge DOS at $\varepsilon=0$ (see Fig.~\ref{fig:QWZmodel}(b)). For $\mu=0$ and $\mu=2t$, the gap closes and both the bulk DOS and edge DOS cancel at $\varepsilon=0$ (Fig.~\ref{fig:QWZmodel}(a) and (c)). The closing of the gap in the DOS leads to a topological phase transition, as can be seen in Fig.~\ref{fig:QWZmodel}(d) where a change in the Chern number~$\mathcal{C}$ occurs: $\mathcal{C}$ is equal to $-1$ in the interval $[\,-2t\,,\,0\,]$, $+1$ in the interval $[\,0\,,\,2t\,]$ and 0 otherwise\cite{Estake2025}. Note that the values $\mu=0$ and $\mu=\pm 2t$ are the only possible values for which one can simultaneously have $\mathcal{E}_{k_xk_y}=|\mathcal{F}_{k_xk_y}|=0$. The edge spectral function $A_{k_xy}(\varepsilon)$ at $y=a$ is shown in Fig.~\ref{fig:QWZspectral1} for various values of $\mu$ in the interval $[\,0\,,\,3t\,[$. At $\mu=0$ and $\mu=2t$ there is a gap closing in the spectral function, in line with the fact that the Chern number value changes for these values of $\mu$, as seen in Fig.~\ref{fig:QWZmodel}(d).  For $\mu\in[\,0\,,\,2t\,[$, it exhibits one chiral edge state within the gap, in agreement with the value of the Chern number $\mathcal{C}$ equal to 1 in that interval. We note that $\mathcal{C}$ does not depend on $\Delta$ in the QWZ model. For $\mu>2t$, no edge state is visible in the spectral function, meaning that we are in the trivial state in that case, consistent with the fact that $\mathcal{C}=0$.

\begin{figure}[t]
\begin{tikzpicture}
\node at (0,0) {
    \includegraphics[height=0.45\linewidth]{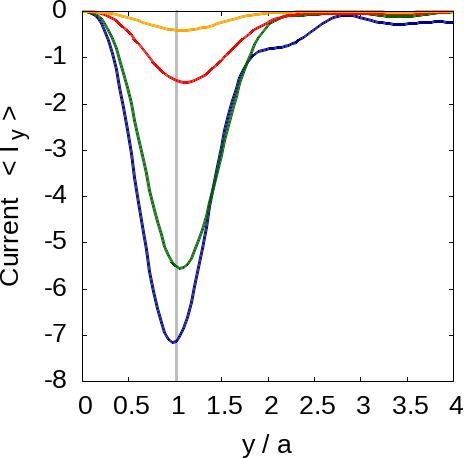}
    \includegraphics[height=0.45\linewidth]{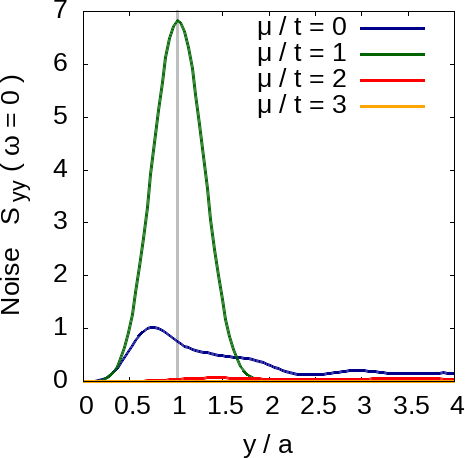}};
\node at (-4.1,1.6) {(a)};
\node at (0.2,1.6) {(b)};
\end{tikzpicture}
    \caption{(a) Current $\langle I_y\rangle$ and (b) zero-frequency noise $\mathcal{S}_{yy}(\omega=0)$ in the QWZ model as a function of $y$, for various values of~$\mu$, at $k_BT=0.1t$, $\Delta=t$, $\eta= 0.02t$ and $U_0=1000t$. }
    \label{fig:QWZfcty}
\end{figure}

The emergence of an ingap chiral edge state enables the flow of electrical charges along the edge of the sample, resulting in the presence of a current and its fluctuations (noise), as can be seen in Fig.~\ref{fig:QWZfcty} where both current  $\langle I_y\rangle$ and zero-frequency noise $\mathcal{S}_{yy}(\omega=0)$ are plotted as a function of the distance $y$ to the impurity line modeling the edge of the sample. We observe that the current is extremum close to $y=a$, and decreases in absolute value with increasing $\mu$. In contrast, zero-frequency noise behaves in a non-monotonic way with increasing $\mu$. Both quantities, current and noise,  decrease towards zero with increasing $y$.

\begin{figure}[t]
\begin{tikzpicture}
\node at (0,0) {
    \includegraphics[height=0.44\linewidth]{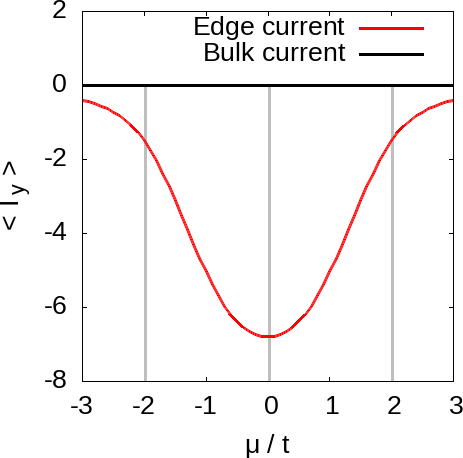}
    \includegraphics[height=0.44\linewidth]{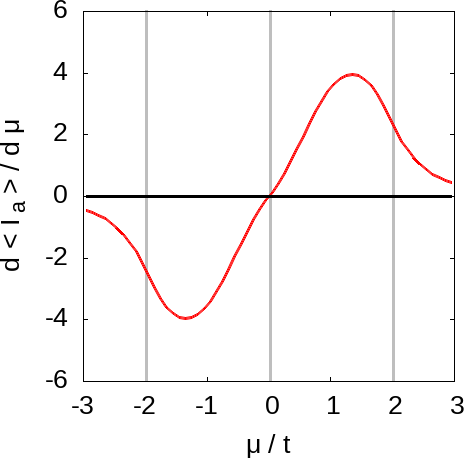}};
\node at (-3.9,1.7) {(a)};
\node at (0.2,1.7) {(b)};
\end{tikzpicture}
\begin{tikzpicture}
\node at (0,0) {
    \includegraphics[height=0.44\linewidth]{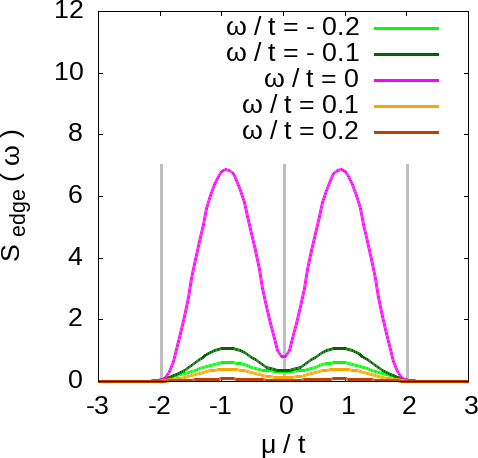}
    \includegraphics[height=0.44\linewidth]{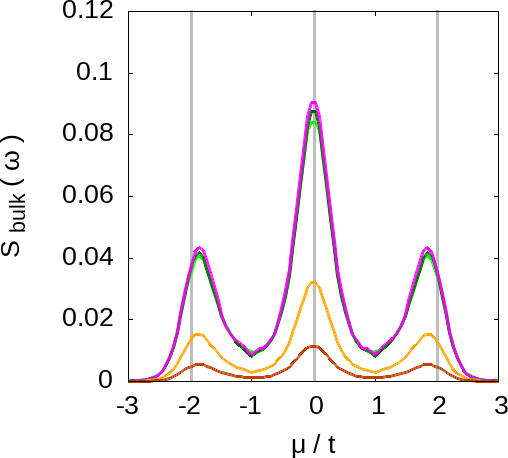}};
\node at (-3.9,1.6) {(c)};
\node at (0.2,1.6) {(d)};
\end{tikzpicture}
    \caption{(a) Edge current $\langle  I_a\rangle$ and bulk current $\langle  I_{y\gg a}\rangle$, (b)~derivative of the edge current according to $\mu$, (c) edge noise and (d) bulk noise in the QWZ model as a function of $\mu$, at $k_BT=0.1t$, $\Delta=t$, $\eta= 0.02t$ and $U_0=1000t$. In~(c) and (d), the noise is plotted for various values of frequency~$\omega$. The vertical gray lines indicate the $\mu$ values for which the Chern number changes.}
    \label{fig:QWZcurrentnoise}
\end{figure}

 In order to further explore the $\mu$-dependence, we plot the current and noise as a function of $\mu$ in Fig.~\ref{fig:QWZcurrentnoise}, at position $y=a$ (edge current and noise) and position $y\gg a$ (bulk current and noise). With regard to the current, Fig.~\ref{fig:QWZcurrentnoise}(a) shows that the bulk current is equal to zero, as expected since there is a gap with no ingap state in that case, whereas the edge current is non-zero. By calculating the derivative of the edge current $\langle I_a\rangle$ according to~$\mu$, it appears that the general profile of $d\langle I_a\rangle/d\mu$, shown in Fig.~\ref{fig:QWZcurrentnoise}(b),  roughly follows the Chern number dependence on $\mu$ (see Fig.~\ref{fig:QWZmodel}(d)).
 

\begin{figure}[b]
\begin{tikzpicture}
\node at (0,0) {
    \includegraphics[height=0.45\linewidth]{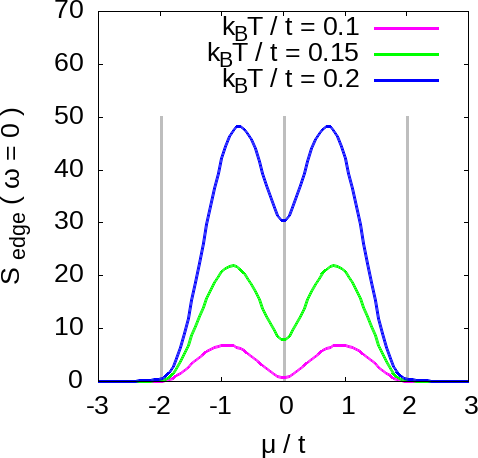}
    \includegraphics[height=0.45\linewidth]{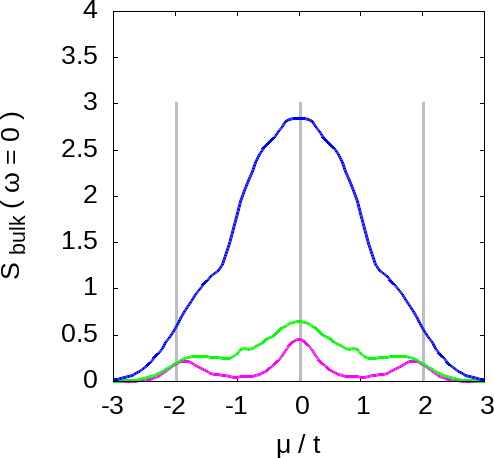}};
\node at (-4.1,1.6) {(a)};
\node at (0.2,1.6) {(b)};
\end{tikzpicture}
    \caption{(a) Edge and (b) bulk noises in the QWZ model as a function of $\mu$, for various values of $k_BT$ at $\Delta=t$, $\eta= 0.02t$ and $U_0=1000t$. }
    \label{fig:QWZtemp}
\end{figure}

We now turn our attention to the results obtained for noise.  Figures~\ref{fig:QWZcurrentnoise}(c), \ref{fig:QWZcurrentnoise}(d), \ref{fig:QWZtemp}(a) and \ref{fig:QWZtemp}(b) give edge and bulk noises as a function of~$\mu$. We observe several notable characteristics: (i)~contrary to bulk current which is equal to zero, bulk noise is non-zero; (ii) bulk noise is two orders of magnitude smaller than edge noise for the set of parameters we use; (iii)~bulk noise exhibits peaks at positions $\mu$ for which the Chern number changes its value, meaning that it behaves like a kind of topological susceptibility; (iv)~edge noise is non-zero only when Chern number is non-zero. These characteristics can be physically justified. Property (i) results from the thermal contribution to the noise. Indeed, since the bulk noise is calculated at finite temperature, it does not vanish like the bulk current does. Nevertheless, since ingap edge states contribute to edge noise but not to bulk noise, the latter  is expected to be negligible compared to the former. This explains property (ii). With regard to (iii), since the Chern number is a topological invariant, it is a bulk property. It is therefore not surprising that a link between the Chern number and bulk noise exists. By looking to the frequency and temperature dependence of the edge and bulk noises, displayed in Figs.~\ref{fig:QWZcurrentnoise} and \ref{fig:QWZtemp}, one observes that the noise decreases when the frequency $\omega$ increases while it increases when the temperature $T$ increases. It can be understood through the Johnson-Nyquist relation at equilibrium which reads $S(\omega)=2\hbar\omega N(\omega)G_\text{ac}(\omega)$, where $G_\text{ac}(\omega)$ is the ac-conductance and $N(\omega)$, the Bose-Einstein distribution function. At constant temperature, this relation explains why the noise decreases when frequency increases. At zero frequency, it transforms into $S(0)=2k_BTG_\text{dc}$, explaining why the noise increases with temperature. Note that it is a power law dependence with $T$ rather than a linear one because the dc-conductance $G_\text{dc}$ also varies with $T$. The detailed structure of the bulk noise at very low temperature, i.e. for $k_BT<0.1 t$, and the relation between the derivative of the edge current and the Chern number will be  discussed in Sec.~\ref{summary}.

In summary, in equilibrium, the bulk noise can be seen as a kind of topological susceptibility, which exhibits peaks every time the Chern number undergoes a value change, whereas the edge noise is related to the conductance, itself related to the Chern number. The peak height in the bulk noise is proportional to the derivative of the Chern number according to $\mu$. It explains why the height of the bulk noise peak at $\mu=0$ is equal to two times the height of the bulk noise peaks at $\mu=\pm 2t$ (see Fig.~\ref{fig:QWZcurrentnoise}(d)).


\begin{figure}[t]
\begin{tikzpicture}
\node at (0,0) {
    \includegraphics[height=0.44\linewidth]{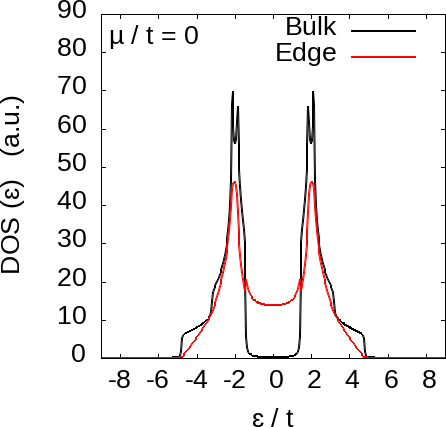}
    \includegraphics[height=0.44\linewidth]{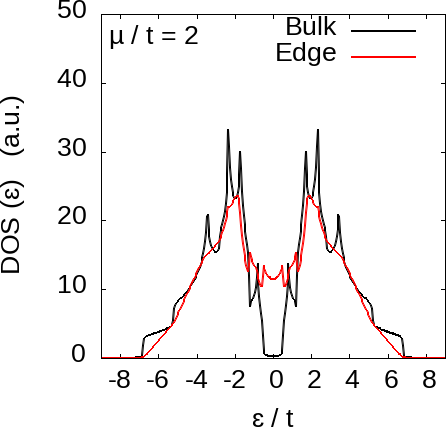}
    };
\node at (-3.8,1.6) {(a)};
\node at (0.3,1.6) {(b)};
\end{tikzpicture}
\begin{tikzpicture}
\node at (0,0) {   
    \includegraphics[height=0.44\linewidth]{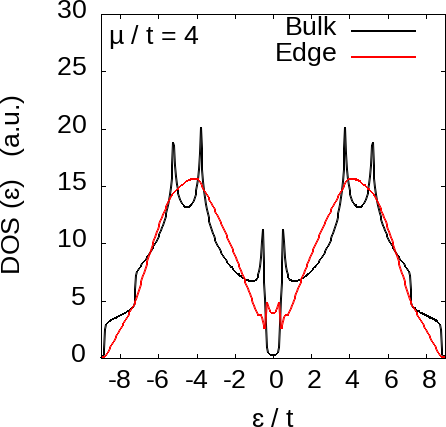}
    \includegraphics[height=0.44\linewidth]{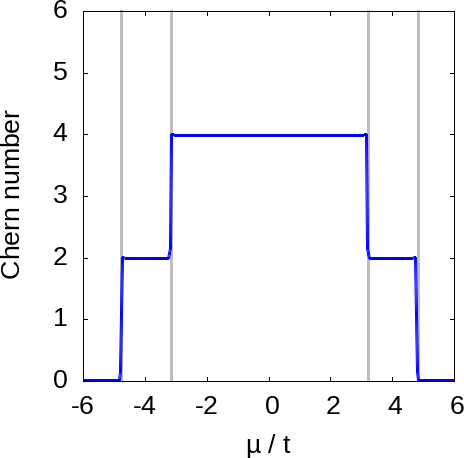}
    };
\node at (-3.8,1.6) {(c)};
\node at (0.3,1.6) {(d)};
\end{tikzpicture}
    \caption{Bulk and edge DOS at (a)~$\mu=0$, (b)~$\mu=2t$, (c)~$\mu=4t$, (d) and Chern number value as a function of $\mu$ in the bilayer model. The parameters are $\Delta=t$, $\gamma=0.8t$, $\eta= 0.02t$ and $U_0=1000t$. In panel (d), the vertical gray lines indicate the values of $\mu$ for which the Chern number value changes, i.e. $\mu=\pm 3.2\,t$ and $\mu=\pm 4.8\,t$.}
    \label{fig:PATHAKmodel}
\end{figure}

\begin{figure}[t]
    \centering
    \includegraphics[width=0.32\linewidth]{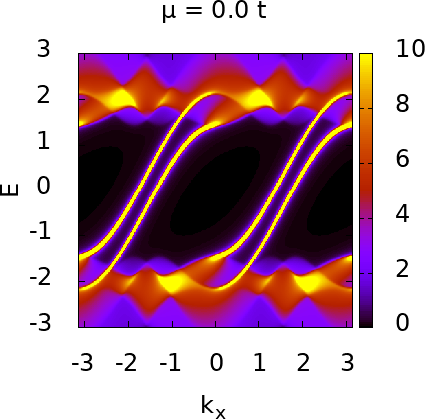}
    \includegraphics[width=0.32\linewidth]{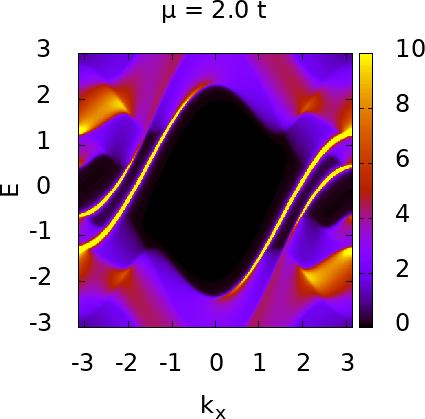} 
    \includegraphics[width=0.32\linewidth]{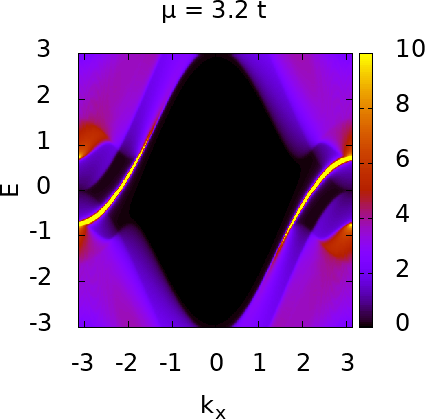} \\
    \vspace{0.2cm} 
    \includegraphics[width=0.32\linewidth]{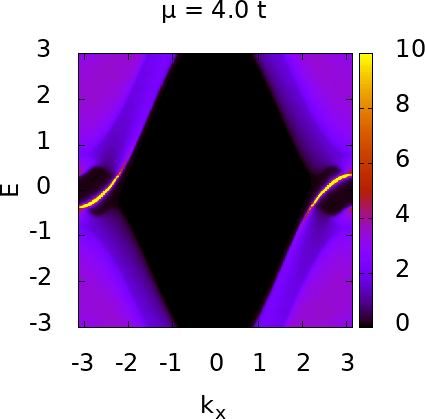}   
    \includegraphics[width=0.32\linewidth]{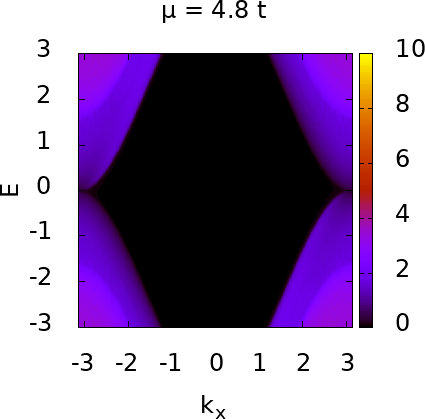}  
    \includegraphics[width=0.32\linewidth]{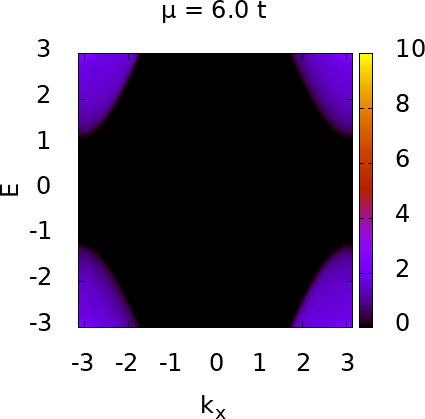}
   \caption{Edge spectral function $A_{k_xy}(\varepsilon)$ as a function of $k_x$ and $\varepsilon$ in the bilayer model at $y=a$ for various values of $\mu$, with $\Delta=t$, $\gamma=0.8t$, $\eta= 0.02t$ and $U_0=1000t$.}
    \label{fig:PATHAKspectral1}
\end{figure}


\subsection{Bilayer model}

Bilayer systems can be described by an effective $4\times 4$ matrix Hamiltonian\cite{Pathak2024} given, in the basis $\{\widehat c_{k_xk_y1},\widehat c_{k_xk_y2},\widehat c^{\,\dagger}_{-k_x-k_y1},\widehat c^{\,\dagger}_{-k_x-k_y2}\}$ where 1 and 2 are the layer indices, by
\begin{eqnarray}
    \mathcal{H}_{k_xk_y}=
    \begin{pmatrix}
    \mathcal{E}_{k_xk_y}& \gamma & \mathcal{F}_{k_xk_y} & 0\\
    \gamma &\mathcal{E}_{k_xk_y} & 0 & \mathcal{F}_{k_xk_y} \\
    \mathcal{F}_{k_xk_y}^* & 0 & -\mathcal{E}_{k_xk_y}& -\gamma\\
    0 & \mathcal{F}_{k_xk_y}^* & -\gamma & -\mathcal{E}_{k_xk_y} \\
    \end{pmatrix}
\end{eqnarray}
with $ \mathcal{E}_{k_xk_y}=-\mu-2t(\cos(ak_x)+\cos(ak_y))$ and $\mathcal{F}_{k_xk_y}=\Delta(\cos(ak_x)-\cos(ak_y)+i\sin(ak_x)\sin(ak_y))$ in the case of $d+id$ superconductivity. $\gamma$ is the inter-layer coupling, $\Delta$~is the superconducting order parameter, and $\mu$ is the chemical potential.  The eigenenergies associated with the clean bilayer system read
\begin{eqnarray}
    E_{\pm\pm}(k_x,k_y)=\pm \sqrt{\left(\mathcal{E}_{k_xk_y}\pm\gamma\right)^{\,2}+|\mathcal{F}_{k_xk_y}|^{\,2}}
\end{eqnarray} 
The impurity matrix~${\bf U}$ used in the T-matrix formalism is taken as
\begin{eqnarray}
   {\bf U}= \begin{pmatrix}
        U_0 \;{\bf I}_2 & 0\\
        0 & -U_0 \;{\bf I}_2
    \end{pmatrix}
\end{eqnarray}
where ${\bf I}_2$ is the $2\times 2$ identity matrix, and $U_0$ is the impurity strength.
The bulk and edge DOSs are given in Fig.~\ref{fig:PATHAKmodel} for $\mu=0$, $\mu=2t$ and $\mu=4t$. Ingap edge states are present at $\varepsilon=0$  (see the red curves). To have a gap closing in the DOS, and a topological phase transition, one must have $(\mathcal{E}_{k_xk_y}\pm\gamma)=|\mathcal{F}_{k_xk_y}|=0$. This is possible only when $\mu=\pm 4t\pm\gamma$. The value chosen for the inter-layer coupling, i.e., $\gamma=0.8t$, gives phase transitions at $\mu=\pm 3.2\,t$ and $\mu=\pm 4.8\,t$. The Chern number is displayed in Fig.~\ref{fig:PATHAKmodel}(d). It takes the value 4 for~$\mu$ in the interval $[\,-3.2\,t\,,\,3.2\,t\,]$, the value 2 for $\mu$ in the intervals $[\,-4.8\,t\,,\,-3.2\,t\,]$ and $[\,3.2\,t\,,\,4.8\,t\,]$, and the value 0 otherwise. It does not depend on $\Delta$.

The edge spectral function $A_{k_xy}(\varepsilon)$ at $y=a$ is shown in Fig.~\ref{fig:PATHAKspectral1} for various values of $\mu$ in the interval $[\,0\,,\,6t\,]$. For $\mu=0$ and $\mu=2 t$, there are four chiral ingap edge states, whereas there are only two chiral ingap edge states for $\mu=3.2 t$ and $\mu=4 t$, in agreement with the Chern number value. For $\mu=4.8 t$ and $\mu=6 t$, there is no ingap state because the Chern number is zero. It confirms the perfect adequacy between the number of ingap edge states and the Chern number. The system undergoes a topological phase transition at $\mu=3.2 t$ and $\mu=4.8 t$, with a gap closing and a change in its Chern number value.

\begin{figure}[t]
\begin{tikzpicture}
\node at (0,0) {
    \includegraphics[height=0.44\linewidth]{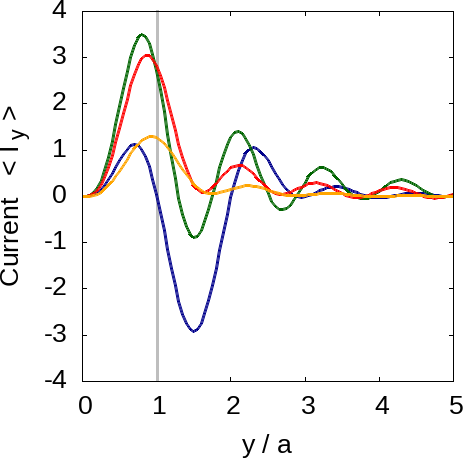}
    \includegraphics[height=0.44\linewidth]{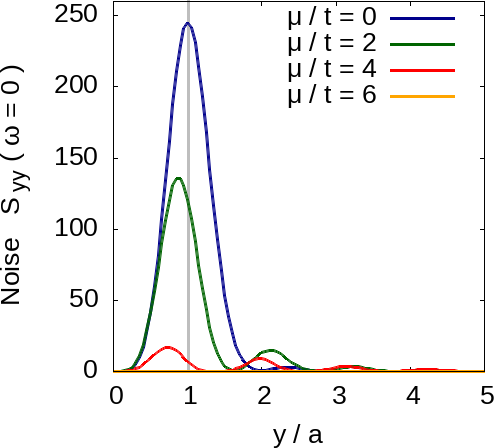}};
\node at (-3.9,1.7) {(a)};
\node at (0.1,1.7) {(b)};
\end{tikzpicture}
    \caption{(a) Current $\langle I_y\rangle$ and (b) zero-frequency noise $\mathcal{S}_{yy}(\omega=0)$ in the bilayer model as a function of $y$, for various values of $\mu$, at $k_BT=0.1t$, $\Delta=t$, $\gamma=0.8t$, $\eta= 0.02t$ and $U_0=1000t$. }
    \label{fig:PATHAKfcty}
\end{figure}

\begin{figure}[t]
\begin{tikzpicture}
\node at (0,0) {
    \includegraphics[height=0.44\linewidth]{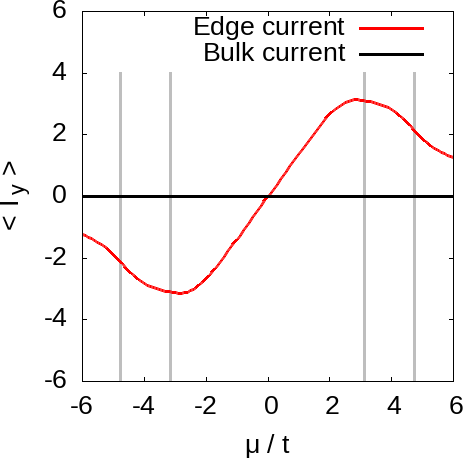}
    \includegraphics[height=0.44\linewidth]{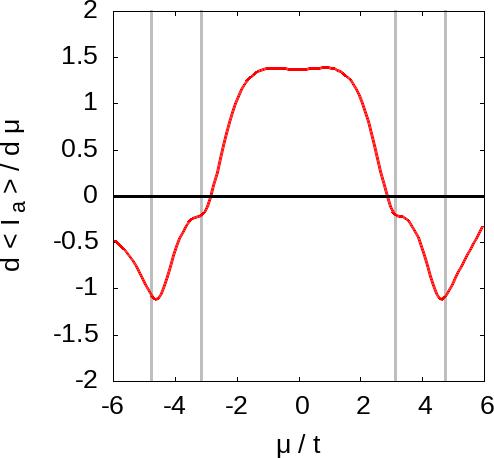}};
\node at (-4,1.6) {(a)};
\node at (0.2,1.6) {(b)};
\end{tikzpicture}
\begin{tikzpicture}
\node at (0,0) {
    \includegraphics[height=0.44\linewidth]{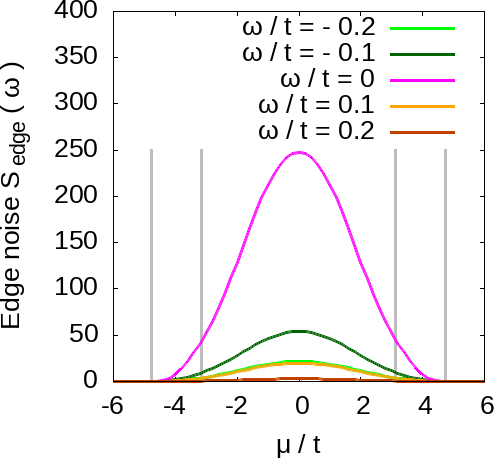}
    \includegraphics[height=0.44\linewidth]{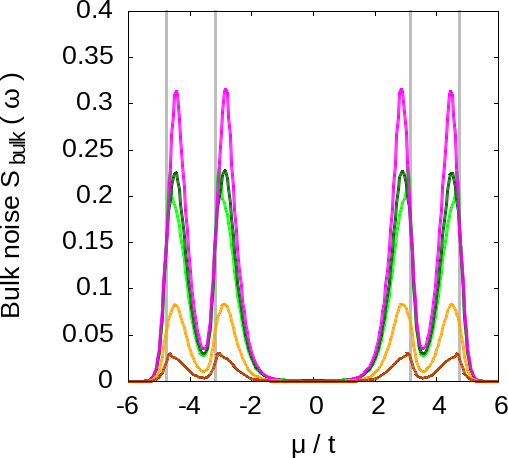}};
\node at (-4,1.6) {(c)};
\node at (0.2,1.6) {(d)};
\end{tikzpicture}
    \caption{(a) Edge current $\langle  I_a\rangle$ and bulk current $\langle  I_{y\gg a}\rangle$, (b)~the derivative of the edge current, (c) edge noise and (d) bulk noise in the bilayer model as a function of $\mu$, at $k_BT=0.1t$, $\Delta=t$, $\gamma=0.8t$, $\eta= 0.02t$ and $U_0=1000t$. In (c) and~(d), the noise is plotted for various values of frequency $\omega$. The vertical gray lines indicate the $\mu$ values for which the Chern number changes.}
    \label{fig:PATHAKcurrentnoise}
\end{figure}

The current and noise dependences according to the distance $y$ are plotted in Fig.~\ref{fig:PATHAKfcty} for various values of $\mu$. We observe damped oscillation behavior in current and noise. They are both extrema in the proximity of position $y=a$, except the current at $\mu=0$ which shows large oscillations.
Figures \ref{fig:PATHAKcurrentnoise}(a) and~(b) show the edge and bulk currents $\langle  I_a\rangle$  and the edge current derivative $\text{d}\langle  I_a\rangle/\text{d}\mu$  as a function of $\mu$. The bulk current is strictly zero, as expected since there is no ingap state in that case. The edge current changes sign at $\mu=0$, which means that the direction of charge propagation on the edge state can be reversed by changing the value of $\mu$. Furthermore, the derivative of the edge current according to $\mu$ shows a staircase behavior, like the Chern number does (see Fig.~\ref{fig:PATHAKmodel}(d)), with steps at the same values for~$\mu$ as for the Chern number, i.e. $\mu=3.2 t$ and $\mu=4.8 t$. This confirms once again the existence of a close relationship between the current derivative and the Chern number. Figures \ref{fig:PATHAKcurrentnoise}(c) and (d) show the edge and bulk noises as a function of $\mu$ at various frequencies. All the characteristics obtained for the noise in the QWZ model are also obtained in the bilayer model: bulk noise is non-zero, although two orders of magnitude smaller than edge noise, and exhibits peaks at $\mu$ values where the Chern number changes; edge noise is non-zero only in the interval where the Chern number is non-zero. In contrast to the current which changes sign at $\mu=0$, the noise always remains positive because we  consider auto-correlator noise $\mathcal{S}_{aa}(\omega)$ in Fig.~\ref{fig:PATHAKcurrentnoise}.

\begin{figure}[t]
\begin{tikzpicture}
\node at (0,0) {
    \includegraphics[height=0.44\linewidth]{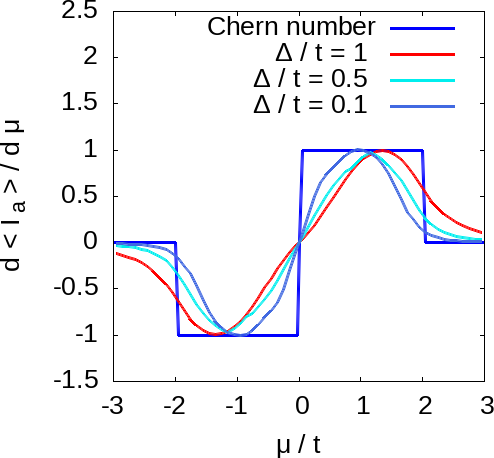}
    \includegraphics[height=0.44\linewidth]{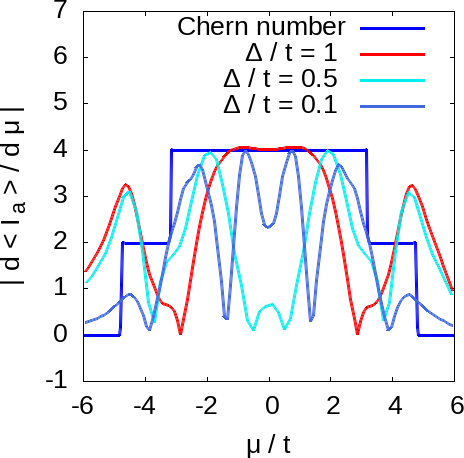}};
\node at (-3.8,1.7) {(a)};
\node at (0.3,1.7) {(b)};
\end{tikzpicture}
    \caption{Derivative of the edge current in the (a) QWZ model, and (b) bilayer model as a function of $\mu$, for various values of $\Delta$ at $k_BT=0.1t$, $\gamma=0.8t$, and $U_0=1000t$. To allow a comparison, all the curves have been vertically rescaled, and, in the case of the bilayer model, the absolute value of the current derivative is taken. The Chern number is also shown.}
    \label{fig:currentDelta}
\end{figure}

\begin{figure}[t]
\begin{tikzpicture}
\node at (0,0) {
    \includegraphics[height=0.44\linewidth]{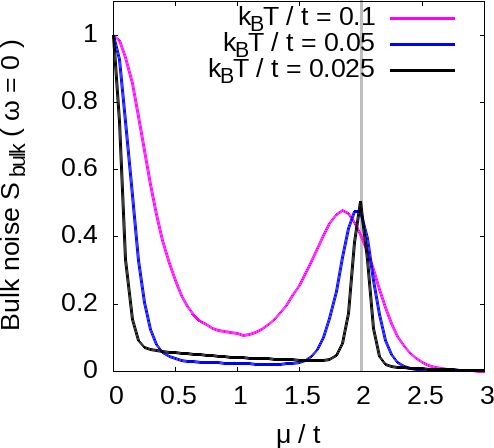}
    \includegraphics[height=0.44\linewidth]{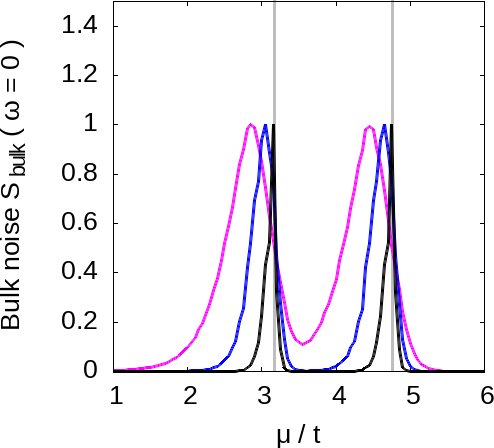}};
\node at (-3.8,1.75) {(a)};
\node at (0.2,1.75) {(b)};
\end{tikzpicture}
    \caption{Bulk noise in the (a) QWZ model, and (b)  bilayer model, as a function of $\mu$, at $\omega=0$, $\Delta=t$, $\gamma=0.8t$, and $U_0=1000t$. We take $\eta= 0.02t$ for $k_BT=0.1t$, $\eta= 0.01t$ for $k_BT=0.05t$ and $\eta= 0.005t$ for $k_BT=0.025t$. The vertical gray lines  indicate the $\mu$ values for which the Chern number changes its value. To allow a comparison, all the curves have been vertically rescaled.}
    \label{fig:noiseT}
\end{figure}

\subsection{Discussion}\label{summary}

By comparing the results obtained in the two previous sections, using the QWZ model and using the bilayer model, we observe similar behaviors in bulk and edge currents and noises, all related to the Chern number, as summarized in Tab.~\ref{tab:synthesis}. Additional details of the comparison are given in the following.

If the absolute value of the derivative of the edge current is proportional to the Chern number, it should cancel out when the Chern number cancels out. However, that is not what we see in Figs.~\ref{fig:QWZcurrentnoise}(b) and~\ref{fig:PATHAKcurrentnoise}(b) which suggests that the current is non-zero even though there is no ingap edge state. This is related to the fact that one has $k_x$-asymmetry in the edge spectral function that gives a contribution to the current, even if there is no ingap state. No such thing occurs for edge noise due to the $\sin^2(k_xa)$ factor in its expression (see Eq.~(\ref{result_noise_equilibrium})). To consolidate our justification, we plot in Fig.~\ref{fig:currentDelta} the edge current derivative for various values of the superconducting parameter $\Delta$ for both the QWZ and bilayer models. This shows that when we reduce the $k_x$-asymmetry by reducing $\Delta$, the profile of the absolute value of the edge current derivative increasingly tends towards the absolute value of the Chern number $\mathcal{C}$. A derivation of the approximate relation $|\text{d}\langle  I_y \rangle/\text{d}\mu|\sim|\mathcal{C}|$ starting from Eqs.~(\ref{result_current}) and (\ref{chern_def}) is given in Appendix \ref{dIC}. In the bilayer model, however, we note the presence of dips in the edge current derivative for some $\mu$ values that are not present in the Chern number curves.

A careful observation of Figs.~\ref{fig:QWZcurrentnoise}(d) and~\ref{fig:PATHAKcurrentnoise}(d) shows that the peaks in the bulk noise are not exactly located at the $\mu$ values for which the Chern number $\mathcal{C}$ changes: there is a slight shift between the peaks and the expected $\mu$ values (highlighted by the vertical gray lines). This is related to the finite value of the temperature. To confirm this, we calculated the bulk noise for decreasing temperature, and plot the results in Fig.~\ref{fig:noiseT} together with the previous results obtained at $k_BT=0.1 t$. This shows that for $k_BT\ll t$, the peak positions in the bulk noise coincide with the positions at which the Chern number changes.

\begin{table}
    \centering
    \begin{tabular}{|l||c|c|}
    \hline
        &  Edge & Bulk \\
       \hline\hline
   Current $\langle  I_y\rangle$     &  $\left|\cfrac{\text{d}\langle  I_a \rangle}{\text{d}\mu}\right|\sim |\mathcal{C}|$ when $\Delta \ll t $  &  0 \\
          \hline
    Noise $\mathcal{S}_{yy}(\omega)\;$      &
    $\;\mathcal{S}_\text{edge}(\omega)\ne 0$ when $\mathcal{C}\ne 0\;$ & $\;\mathcal{S}_\text{bulk}(\omega)\sim \left|\cfrac{d\mathcal{C}}{d\mu}\right|\;$\\
    & $\;\mathcal{S}_\text{edge}(\omega)= 0$ when $\mathcal{C}= 0\;$ & when $k_BT\ll t$ \\
    \hline
    \end{tabular}
    \caption{Synthesis of the results obtained for bulk and edge currents and noises, in both the QWZ and bilayer models.}
    \label{tab:synthesis}
\end{table}

In Appendix \ref{appA}, we extend our study to an anisotropic model for which a nontrivial topological phase can exist even though the Chern number is zero, and we show that edge current and noise take non-zero value in the associated region, meaning that they are related to other types of topological invariant, namely the Zak phase in that case.



\section{Conclusion}

The current fluctuations in a 2D topological superconductors were calculated and expressed as an integral over energy and momentum involving the spectral function and the Fermi-Dirac distribution function. The presence of a $\sin^2(k_xa)$ factor leads to the possibility of non-zero current fluctuations for both chiral and non-chiral edge states, even though the current is zero in the latter case due to the $\sin(k_xa)$ factor in its expression. We then studied the noise behavior in two different models exhibiting chiral edge states, i.e. the QWZ model and bilayer model, and highlight interesting similarities between the results. The edge noise is non-zero when the Chern number $\mathcal{C}$ is non-zero, whereas the bulk noise behaves like the absolute value of the derivative of $\mathcal{C}$ according to the chemical potential. This result means that the latter quantity, the bulk noise, could be seen as a topological susceptibility with a peak each time that $\mathcal{C}$ changes value. The absolute value of the edge current derivative according to the chemical potential is also closely related to the Chern number, although additional dips are obtained in the former quantity for the bilayer model. The study of noise in more realistic systems, and for other types of superconducting pairing, such as $s+id$ pairing, is left to future work with the aim of confirming or refuting the universality of these results.


\begin{acknowledgements}
A.C. thanks C. Bena, E. Pangburn and C. P\'epin for valuable discussions, and MesoNET for the allocation of HPC numerical resources through Project No. M23023 supported by a French government grant managed by the Agence Nationale de la Recherche under the Investissements d’avenir program (ANR-21-ESRE-0051).
\end{acknowledgements}


\appendix

\section{Relation between the Chern number and the derivative of the edge current}\label{dIC}

From Eq.~(\ref{chern_def}), we can see that the Chern number is given by integrals of the following form 
\begin{eqnarray}
 \mathcal{C}&\sim&\iiint \text{d}{k_x}\text{d}{k_y} \text{d}\varepsilon(\partial_{k_y}\mathcal{H})G(\varepsilon)(\partial_{k_x}\mathcal{H})G^2(\varepsilon)
\end{eqnarray}
An estimation gives $ \mathcal{C}
 \sim\iint \text{d}{k_x} \text{d}\varepsilon\mathcal{H}G(\varepsilon)(\partial_{k_x}\mathcal{H})G^2(\varepsilon)\sim -t\iint \text{d}{k_x} \text{d}\varepsilon\sin(ak_x) G^2(\varepsilon)$, where we have used $\mathcal{H}\sim \mu-t(\cos(k_xa)+\cos(k_ya))$ and~$\mathcal{H}G(\varepsilon)=\varepsilon G(\varepsilon)-1$. In the latter relation, the first term does not contribute to the integral over energy since it is an odd function. From Eq.~(\ref{result_current}), we can see that the current is given by an integral of the form
$\langle I_y\rangle\sim t\iint \text{d}k_x \text{d}\varepsilon\sin(ak_x)  G(\varepsilon)$. The derivative of $\langle I_y\rangle$ according to $\mu$ is $
 \text{d} \langle I_y\rangle/\text{d}\mu \sim t\iint \text{d}k_x\text{d}\varepsilon \sin(ak_x)f(\varepsilon) \text{d}G(\varepsilon)/\text{d}\mu$,
where $\text{d}G(\varepsilon)/\text{d}\mu\sim -G^2(\varepsilon)$ since $\text{d}\mathcal{H}/\text{d}\mu=1$, leading to the relation $\text{d} \langle I_y\rangle/\text{d}\mu\sim \mathcal{C}$. Indeed, at a sufficiently low temperature, the Fermi-Dirac distribution function reads $f(\varepsilon)=\Theta(-\varepsilon)$, where $\Theta$ is the Heaviside function.

\begin{figure}[t]
\begin{tikzpicture}
\node at (0,0) {
    \includegraphics[height=0.44\linewidth]{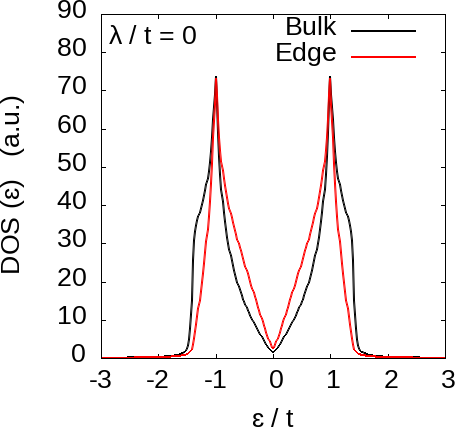}
    \includegraphics[height=0.44\linewidth]{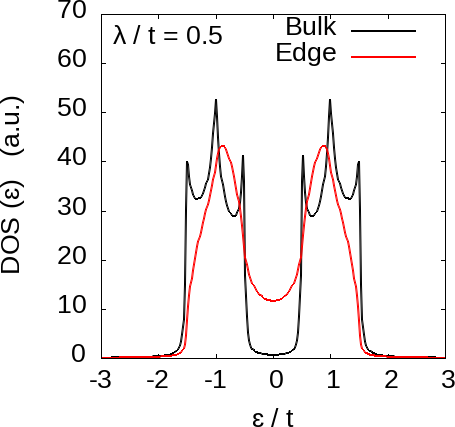}
    };
\node at (-3.9,1.6) {(a)};
\node at (0.3,1.6) {(b)};
\end{tikzpicture}
\begin{tikzpicture}
\node at (0,0) {   
    \includegraphics[height=0.44\linewidth]{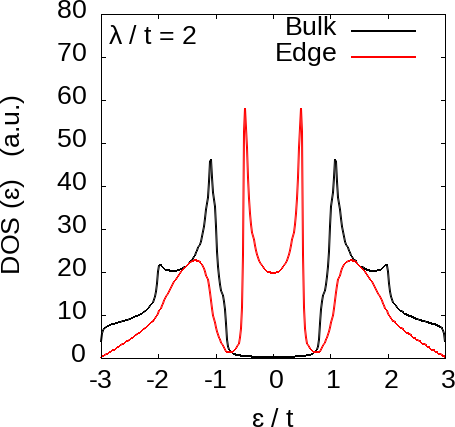}
    \includegraphics[height=0.44\linewidth]{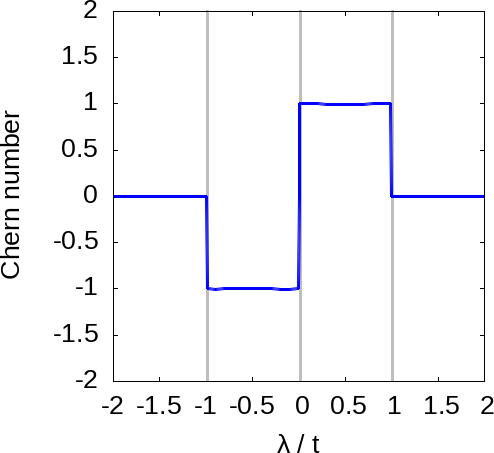}
    };
\node at (-3.9,1.6) {(c)};
\node at (0.3,1.6) {(d)};
\end{tikzpicture}
    \caption{Bulk and edge DOSs at (a)~$\lambda=0$, (b)~$\lambda=t$, (c)~$\lambda=2t$, and (d) the Chern number $\mathcal{C}$ as a function of $\lambda$ in the extended QWZ model. The parameters are $m=0.5 t$, $\eta= 0.02t$ and $U_0=1000t$.  In panel (d), the vertical gray lines indicate the values of $\lambda$ for which the Chern number value changes, i.e. $\lambda=0$ and $\lambda=\pm t$.}
    \label{fig:WUmodel}
\end{figure}

\begin{figure}
    \centering 
    \includegraphics[width=0.4\linewidth]{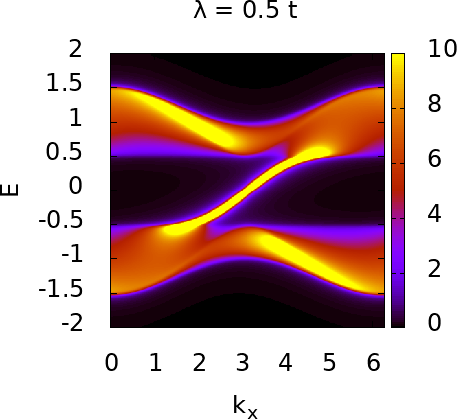}  
    \includegraphics[width=0.4\linewidth]{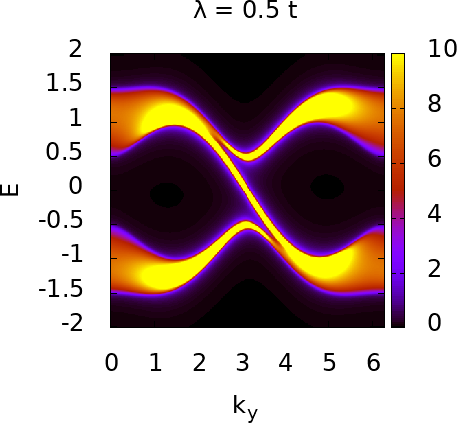}\\
    \vspace{0.2cm}
    \includegraphics[width=0.4\linewidth]{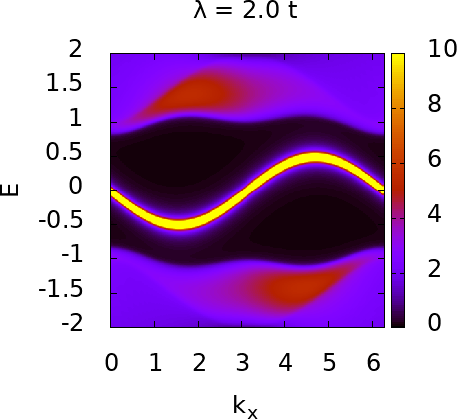}  
    \includegraphics[width=0.4\linewidth]{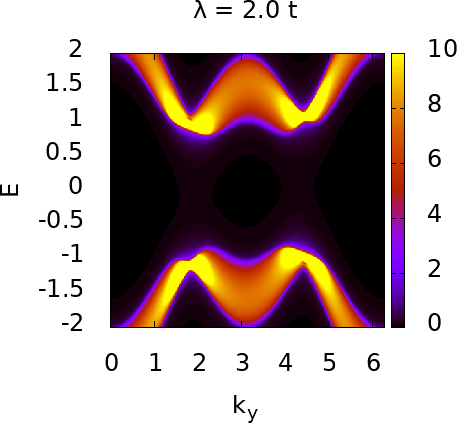}
   \caption{Edge spectral functions $A_{k_xa}(\varepsilon)$ along the $k_x$-axis (left column) and $A_{k_ya}(\varepsilon)$ along the $k_y$-axis (right column), in the extended QWZ model for $\lambda=0.5 t$ and $\lambda=2 t$, with $m=0.5 t$, $\eta= 0.02t$ and $U_0=1000t$.}
    \label{fig:WUspectral}
\end{figure}

\begin{figure}[t]
\begin{tikzpicture}
\node at (0,0) {
    \includegraphics[height=0.44\linewidth]{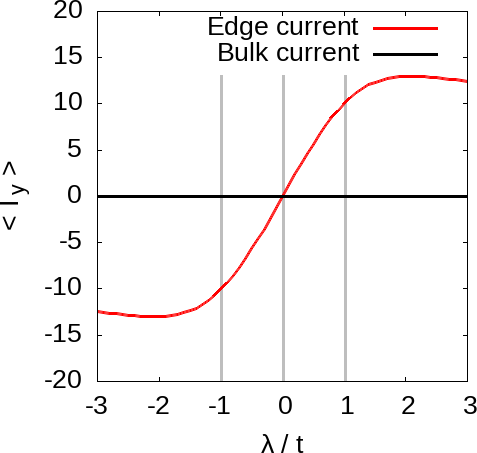}
    \includegraphics[height=0.44\linewidth]{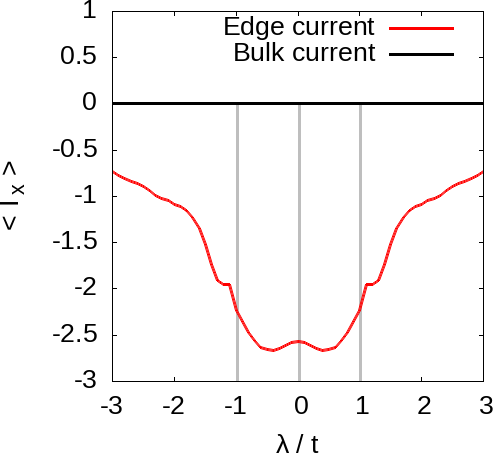}};
\node at (-3.9,1.7) {(a)};
\node at (0.2,1.7) {(b)};
\end{tikzpicture}
\begin{tikzpicture}
\node at (0,0) {
    \includegraphics[height=0.44\linewidth]{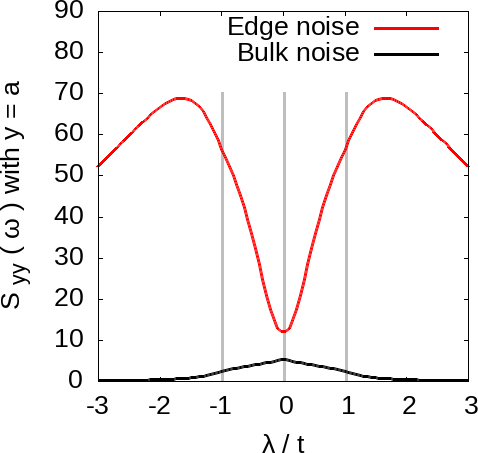}
    \includegraphics[height=0.44\linewidth]{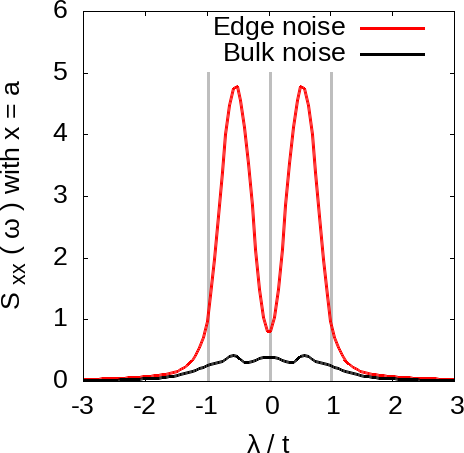}};
\node at (-3.9,1.6) {(c)};
\node at (0.2,1.6) {(d)};
\end{tikzpicture}
    \caption{(a) Edge current $\langle  I_{y=a}\rangle$ and bulk current $\langle  I_{y\gg a}\rangle$, (b)~edge current $\langle  I_{x=a}\rangle$ and bulk current $\langle  I_{x\gg a}\rangle$, (c) edge and bulk noise $S_{yy}(\omega)$, and (d) edge and bulk noise $S_{xx}(\omega)$ in the extended QWZ model as a function of $\lambda$, at $\omega=0$, $k_BT=0.1t$, $m=0.5 t$, $\eta= 0.02t$ and $U_0=1000t$. The vertical gray lines indicate the $\lambda$ values for which the Chern number changes.}
    \label{fig:WUcurrentnoise}
\end{figure}

\section{Anisotropic edges}\label{appA}

To complete the study, we consider an anisotropic model for which the edge states in the $x$-direction and in the $y$-direction have radically different characteristics, and show how current and noise evolve in that case, in particular in the region where the Chern number is zero. To that end, we introduce an extended QWZ model\cite{Wu2020} given by the $2\times 2$ matrix Hamiltonian
\begin{eqnarray}
    \mathcal{H}_{k_xk_y}=
    \begin{pmatrix}
    \mathcal{E}_{k_xk_y}& \mathcal{F}_{k_xk_y}\\
    \mathcal{F}^*_{k_xk_y}&-\mathcal{E}_{k_xk_y}\\
    \end{pmatrix}
\end{eqnarray}
with $\mathcal{E}_{k_xk_y}=t\sin(ak_y)$ and $ \mathcal{F}_{k_xk_y}=m(1+\exp(-iak_x))+\lambda\cos(a k_y)$, where $t$ and $m$ are first neighbors hopping integrals, and $\lambda$ is related to the effective spin-orbit coupling. Figure~\ref{fig:WUmodel} shows the DOS for various values of $\lambda$, and the evolution of the Chern number as a function of $\lambda$: its takes the value -1 in the interval $\lambda\in[-1,0]$, the value +1 in the interval $\lambda\in[0,1]$, and the value 0 otherwise. For $\lambda=2t$, a value for which the Chern number is zero, we note the presence of strong ingap edge states (see red curve in Fig.~\ref{fig:WUmodel}(c)), related to a nontrivial topological phase with zero Berry curvature but non-zero Berry connection\cite{Liu2017}. It is this phase that we will explore here.

The spectral function $A_{k_xy}(\varepsilon)$, current $\langle I_{y}\rangle$ and noise $S_{yy}(\omega)$, associated with an edge located at $y=a$, are calculated using Eqs.~(\ref{def_spectral_function}), (\ref{result_current_pathak}) and (\ref{result_noise_equilibrium}). The spectral function $A_{k_yx}(\varepsilon)$, current $\langle I_{x}\rangle$ and noise $S_{xx}(\omega)$, associated with an edge located at $x=a$, are calculated by using similar expressions in which $k_x$ and $k_y$, on the one hand, and $y$ and~$x$, on the other hand, are exchanged. Figure \ref{fig:WUspectral} shows the edge spectral function for an edge at $y=a$ (left column) and an edge at $x=a$ (right column). We remark that for $\lambda=2t$, an ingap edge state exists when the edge is located at $y=a$, whereas there is no ingap edge state when the edge is located at $x=a$, which indicates strong anisotropy between the $x$ and $y$ directions for this particular model. We now examine the effect of this anisotropy on transport properties. Figure~\ref{fig:WUcurrentnoise} shows the evolution of current and noise as a function of~$\lambda$, for each of the two edges: edge located at $y=a$ (left column) and edge located at $x=a$ (right column). We remark that in the latter case, the edge current and noise behave in the same way as what we found for the QWZ model (compare Figs.~ \ref{fig:QWZcurrentnoise}(a) and \ref{fig:QWZcurrentnoise}(c) to Figs.~\ref{fig:WUcurrentnoise}(b) and \ref{fig:WUcurrentnoise}(d)). In contrast, for the other edge direction, we observe a radically different behavior with non-canceling edge current and noise at values of~$\lambda$ for which Chern number cancels, with the presence of a persistent edge current and strong  edge noise (see Figs.~\ref{fig:WUcurrentnoise}(a) and \ref{fig:WUcurrentnoise}(c)), that are no longer related to the Chern number, but rather to the projection of wave-polarization, alternatively called Zak phase\cite{Wu2020}. This shows that edge current and noise are not only sensitive to the Chern number value, but could also be related to other topological invariants. This can play a crucial role in cases where the Chern number is no longer the characteristic invariant, which can occur in certain topological regions.


\bibliography{toponoise.bib}

\end{document}